\documentclass[preprint, superscriptaddress, showpacs, amsmath,amssymb]{revtex4}

\usepackage{graphicx}
\begin{document}

% Use the \preprint command to place your local institutional report
% number in the upper righthand corner of the title page in preprint mode.
% Multiple \preprint commands are allowed.
% Use the 'preprintnumbers' class option to override journal defaults
% to display numbers if necessary
%\preprint{}

%Title of paper
\title{Energy cascades and spectra in turbulent Bose-Einstein condensates}
%the defocusing Nonlinear Schr\"odinger  equation}

% repeat the \author .. \affiliation  etc. as needed
% \email, \thanks, \homepage, \altaffiliation all apply to the current
% author. Explanatory text should go in the []'s, actual e-mail
% address or url should go in the {}'s for \email and \homepage.
% Please use the appropriate macro foreach each type of information

% \affiliation command applies to all authors since the last
% \affiliation command. The \affiliation command should follow the
% other information
% \affiliation can be followed by \email, \homepage, \thanks as well.
\author{Davide Proment}
\email{davideproment@gmail.com}
\affiliation{Dipartimento di Fisica Generale, Universit\`{a} di Torino, Via Pietro Giuria 1, 10125 Torino, Italy}

\author{Sergey Nazarenko}
\affiliation{Mathematics Institute, The University of Warwick, Coventry, CV4-7AL, UK}

\author{Miguel Onorato}
\affiliation{Dipartimento di Fisica Generale, Universit\`{a} di Torino, Via Pietro Giuria 1, 10125 Torino, Italy}

%\homepage[]{Your web page}
%\thanks{}
%\altaffiliation{}
\affiliation{}

%Collaboration name if desired (requires use of superscriptaddress
%option in \documentclass). \noaffiliation is required (may also be
%used with the \author command).
%\collaboration can be followed by \email, \homepage, \thanks as well.
%\collaboration{}
%\noaffiliation

\date{\today}

\begin{abstract}
We present a numerical study of turbulence in Bose-Einstein condensates within the 3D Gross-Pitaevskii equation.
We concentrate on the direct energy cascade in forced-dissipated systems.
We show that behavior of the system is very sensitive to the properties of the model at the scales greater than  the forcing scale, and we identify three different universal regimes: (1) a non-stationary regime with condensation and transition from a four-wave to a three-wave interaction process when the largest scales are not dissipated, (2) a steady weak wave turbulence regime when largest scales are dissipated with a friction-type dissipation, (3) a state with a scale-by-scale balance of the linear and the nonlinear timescales when the large-scale dissipation is a hypo-viscosity.
\end{abstract}

% insert suggested PACS numbers in braces on next line
%\pacs{67.85.De, 94.05.Lk, 94.05.Pt, 05.20.Dd}
% insert suggested keywords - APS authors don't need to do this
%\keywords{}

%\maketitle must follow title, authors, abstract, \pacs, and \keywords
\maketitle
Experimental discovery of Bose-Einstein condensates (BEC) in 1995 \cite{anderson1995obe, davis1995bec}, some sixty years after their theoretical prediction \cite{bose1924pgl, einstein1925paw}, sparked a renewed interest in this subject. 
Besides the obvious  importance of such systems for fundamental physics, BEC experiments provide an excellent opportunity to build and study  new nonlinear dynamical systems  fabricated  with high degree of control and flexibility supplied by optical means. 
For the nonlinear science and applied mathematics such an opportunity is extremely valuable, because it allows to implement and test dynamical and statistical regimes previously predicted  theoretically and to gain insights about new ones for which the theory is yet to be developed. 
This is because BEC can be described by one of the most important and universal PDE's, the nonlinear Schr\"odinger equation, called in this case Gross-Pitaevskii equation (GPE) \cite{pitaevskii2003bec}:
\begin{equation}
i\frac{\partial \psi}{\partial t}+\nabla^2\psi - |\psi|^2\psi=
\mathcal{F}+ \mathcal{D},
\label{nls0}
\end{equation}
where $\psi$ is the order parameter indicating the condensate wave function, $\mathcal{F} $ and $ \mathcal{D}$ represent   possible external forcings and dissipation mechanisms. 
In general, when $ \mathcal{F}=0 $ and $ \mathcal{D}=0 $, GPE conserves total energy and particles
\begin{subequations}
\begin{equation}
H = \int \frac{1}{2} |\nabla \psi|^2 d\mathbf{x} + \int \frac{1}{4} |\psi|^4 d\mathbf{x} = H_{LIN} + H_{NL},
\label{H}
\end{equation}
\begin{equation}
N = \int \frac{1}{2} | \psi|^2 d\mathbf{x}.
\label{N}
\end{equation}
\end{subequations}

As GPE model describes a Bose gas at very low temperature, it has been used to study the formation of a condensate in \cite{dyachenko1992wtc, berloff2002ssn, nazarenko2006wta}. 
Moreover GPE can be mapped, using the  Madelung transformation, to the  Euler equation for ideal fluid flows with the extra quantum pressure term. 
This is why many concepts arising from the fluid dynamics have been discussed and studied with GPE, for example vortices and their reconnection \cite{koplik1993vrs}.
It was also suggested that this model allows statistical motions similar to classical fluid turbulence and a number of papers was devoted to finding Kolmogorov spectrum in such GPE turbulence \cite{nore1997dkt, abid2003gpd, parker2005edt, kobayashi2005kss, kobayashi2007qtt}.

On the other hand, GPE solutions also include dispersive waves which may be involved in nonlinear interactions, and an approach known as weak wave turbulence (WWT) can be made for GPE.
Generally, WWT describes statistics on large ensembles of weakly nonlinear waves in different applications, i.e. water waves or waves in plasmas \cite{zakharov41kst}.
Such waves interact with each other in a resonant way, e.g. in triads ar quartets, thereby transferring energy (or/and any other invariants) through the scale space forming turbulent cascades similar to the classical Kolmogorov cascade in hydrodynamic turbulence.
One remarkable property of WWT is that, in contrast to hydrodynamic turbulence, power law spectra corresponding to such cascades, known as Kolmogorov-Zakharov (KZ) spectra, have been found as exact stationary solutions of the corresponding wave kinetic equation \cite{zakharov41kst}.

WWT for GPE turbulence was developed in \cite{zakharov1985had, dyachenko1992wtc}; the following wave kinetic equation was derived
\begin{eqnarray}
\frac{\partial n_1}{\partial t} & = & 4\pi \int n_1 n_2 n_3 n_4 \left(\frac{1}{n_1}+\frac{1}{n_2}-\frac{1}{n_3}-\frac{1}{n_4}\right) \times \label{KIN} \\ 
& & \delta(\mathbf{k}_1+\mathbf{k}_2-\mathbf{k}_3-\mathbf{k}_4) \delta(\omega_1+\omega_2-\omega_3-\omega_4) d\mathbf{k}_{234} \nonumber
\end{eqnarray}
%\begin{eqnarray}
%\frac{\partial n(\mathbf{k})}{\partial t} = 4\pi \int n(\mathbf{k}) n(\mathbf{k}_1) n(\mathbf{k}_2) n(\mathbf{k}_3)  \times  \hspace{2cm}
%\\ \nonumber
%\left[\frac{1}{n(\mathbf{k})}+\frac{1}{n(\mathbf{k}_1)}-\frac{1}{n(\mathbf{k}_2)}-\frac{1}{n(\mathbf{k}_3)}\right]  \delta(\mathbf{k}+\mathbf{k}_1-\mathbf{k}_2-\mathbf{k}_3)
%\\ \nonumber
%\times \delta(\omega(\mathbf{k})+\omega(\mathbf{k}_1) - \omega(\mathbf{k}_2)-\omega(\mathbf{k}_3)) d\mathbf{k}_{1} d\mathbf{k}_{2} d\mathbf{k}_{3}
%\label{KIN}
%\end{eqnarray}
where $ n_i = (L/2\pi)^d  \langle \hat\psi(\mathbf{k}_i, t)  \hat\psi^{\ast}(\mathbf{k}_i, t) \rangle$ is the {\itshape wave-action} spectrum averaged over many realisations (here $L$ is the size of the bounding box and $ d $ is the dimension of the space), $\omega_i=k_i^2$  is the wave frequency, and $ k_i=|\mathbf{k}_i| $. 
Equation (\ref{KIN}) conserves the total wave-action $ N=\int n_1 d\mathbf{k}_1 $ and the total energy $ E=\int \omega_1 n_1 d\mathbf{k}_1 $  which correspond to the GPE invariants (\ref{H}) and (\ref{N}) respectively.
Besides thermodynamic solutions, equation (\ref{KIN}) has two non-equilibrium stationary isotropic solutions of the form of $ n(k) \sim k^{-\alpha} $ corresponding  to  constant fluxes of energy or wave-action (KZ spectra).  
In 3D the {\itshape direct energy cascade} spectrum has $ \alpha=3 $ and the  {\itshape inverse wave-action cascade} has $ \alpha=7/3 $. 
We will present  our  results in terms of the 1D wave-action spectrum $ n^{1D}(k)=4\pi k^2 n(k) $, i.e. after integration  over the solid angle.
For such a spectrum the WWT prediction for the direct cascade is $n^{1D}(k)\sim k^{-1}$.
Note that in hydrodynamic turbulence the results are usually discussed in terms of the 1D energy spectrum  $E^{1D}(k)$ (e.g. Kolmogorov $E^{1D}(k) \sim k^{-5/3}$).
For WWT we have the relation $E^{1D}(k) = \omega({k})  n^{1D}(k)$,  which for GPE means $E^{1D}(k) \sim k^{-\alpha+4}$. 
Further, it was predicted in \cite{dyachenko1992wtc} that in presence of a condensate (due to the inverse wave-action cascade) the four-wave resonant interaction will eventually be replaced by a three-wave process with an acoustic-type KZ spectrum.

Previously, there has been  a number of  numerical simulations of tubulence in 2D GPE case and comparisons with the WWT predictions  \cite{dyachenko1992wtc}, \cite{nazarenko2006wta}. 
For the 3D case, we are aware of a number of simulations of GPE in freely decaying case, \cite{kobayashi2005kss}, or for the unforced, undamped simulation where the initial condition  relaxes to the thermodynamic solution, \cite{berloff2002ssn}. 
As far a we know no steady state (in the sense of cascade) has ever been reached in any simulation, and no direct comparison with the WWT predictions has ever been attempted. 
The purpose of the present work is to revisit the problem of 3D GPE turbulence in the direct energy cascade range using the numerical simulations.
Our goal will be to find the spectrum and to see if (and when) it agrees with the WWT prediction. 
In cases when the numerics yield a spectrum which is different from WWT, we will aim to establish the physical mechanisms behind the formation of this state.
%We will also keep in mind the previous numerical searches for the Kolmogorov spectrum, and we will try to relate our results to these works.

Our numerical domain is a cube with uniform mesh of $ 256^3 $ points and periodic boundary conditions. 
We integrate equation (\ref{nls0}) by a standard split step method. 
In order to observe the cascade,  energy and wave-action are injected directly in Fourier space at wave numbers $ \in [9 \Delta k, 10 \Delta k] $ by a forcing term $ \hat{\mathcal{F}} = -i f_0 e^{i \varphi(k)} $ with $ \varphi $ randomly distributed in $k$-space and time and $f_0$ defining the forcing coefficient.
To absorb energy at high wave numbers and prevent accumulation or thermalization, a dissipative hyper-viscous term $ \mathcal{D}= i \nu_h(- \nabla^{2})^n \psi $ is included in (\ref{nls0}), with $ \nu_h= 2 \times 10^{-6} $ and $ n=8 $.

\begin{figure}
\includegraphics[scale=0.24]{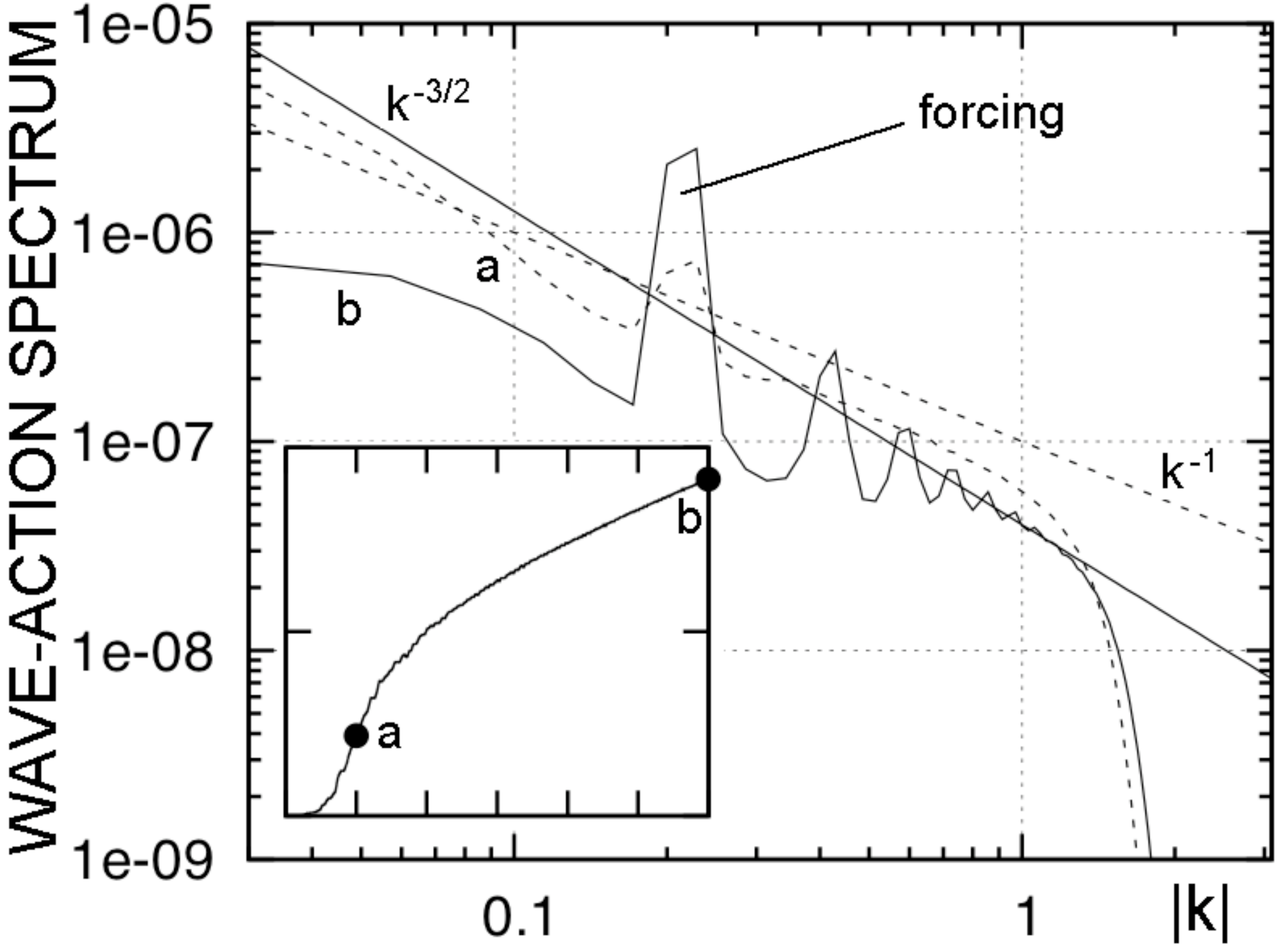}
\caption{Spectrum $ n^{1D}(k, t) $  at $ t=1\times 10^5 $ {\bf (a)} and at $ t=6\times 10^5 $ {\bf (b)}. 
In this simulation the dissipation is only hyper-viscosity. 
Slopes $ k^{-1} $ and  $ k^{-3/2} $ are also plotted. 
Inset: $ c_0=|\hat\psi(0,t)| $ as a function of time.\label{spectrum-hv}}
\end{figure}

If GPE is integrated with a forcing at low wave numbers and a dissipation at only high wave numbers, it will never reach a steady solution. 
This is because it admits also the inverse cascade of wave-action which will start feeding  wave number $ \mathbf{k}=0 $ and its close vicinity, building a strong condensate $ c_0=|\hat\psi(0, t)| $ that changes the type of interactions from four to three-wave resonances \cite{dyachenko1992wtc}. 
This become clear by looking at figure \ref{spectrum-hv} where we show the spectrum at two stages. 
In the first one, {\bf (a)}, the spectrum exhibits at wave numbers larger than forcing a power low close to $k^{-1}$, consistently with the WWT prediction. 
As the simulation evolves, the condensate grows and the spectrum starts to deviate from the pure $-1$ scaling: a set of well-defined peaks appears in the spectrum, curve {\bf (b)}. 
These peaks are probably the result of a three-wave interaction similar to ones previously observed for a three-wave system in \cite{falkovich1988efc}.
In the inset we present the condensate as a function of time: the dots {\bf (a)} and {\bf (b)} correspond to the instant of times at which the two spectra are computed.

\begin{figure}
\includegraphics[scale=0.28]{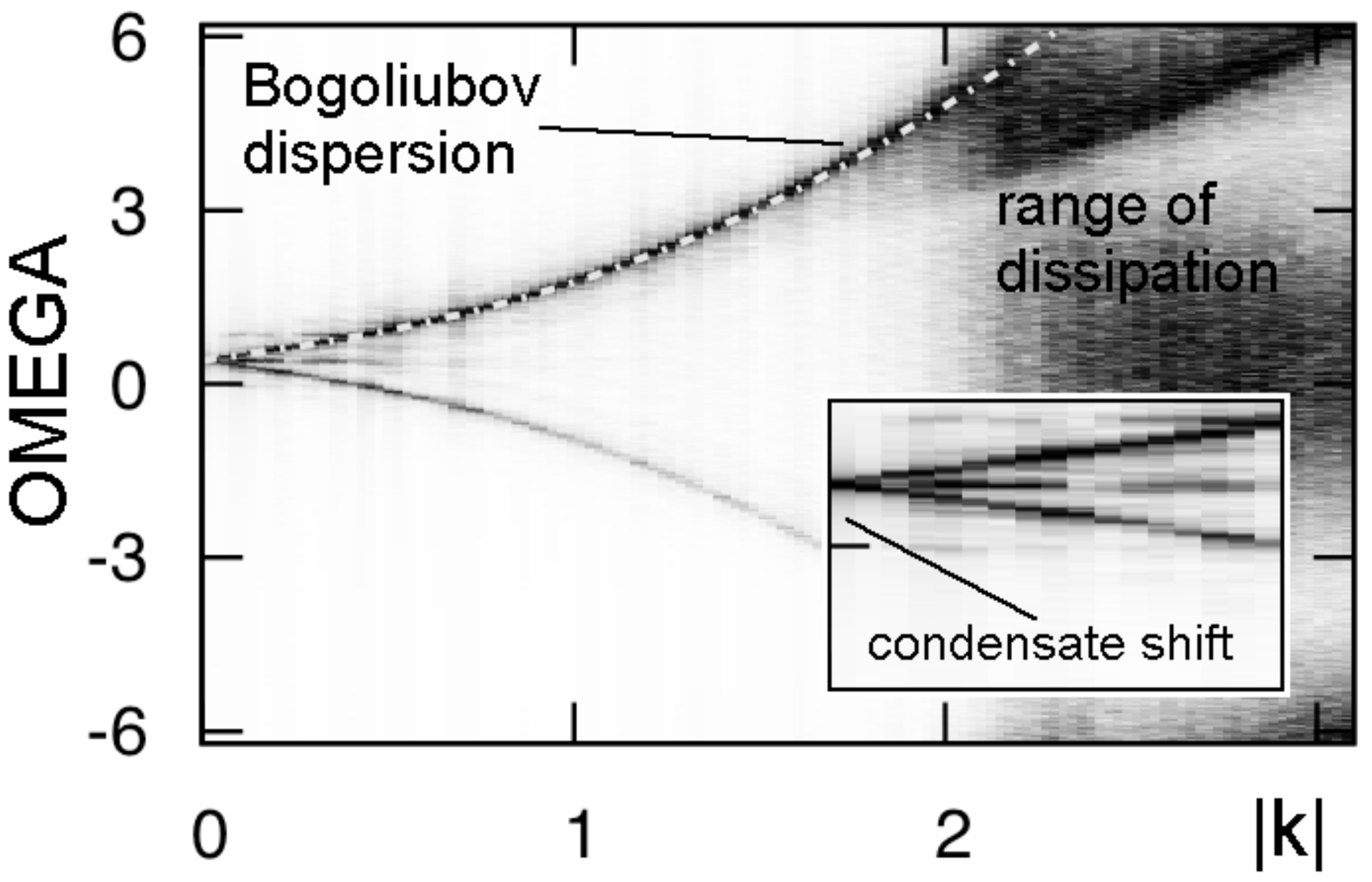}
\caption{Dispersion relation in the simulation with only hyper-viscosity evaluated at final time, see figure \ref{spectrum-hv} case {\bf (b)}. 
The Bogoliubov dispersion curve (only positive branch) is superposed with white dashed line. 
Inset: zoom in the zone of low $ k $'s to observe  the condensate (horizontal branch).\label{dispersion-hv}}
\end{figure}

Regime where the condensate is prevalent was theoretically considered in \cite{dyachenko1992wtc, zakharov2005dbe}. 
In this case, the wave field in (\ref{nls0}) can be decomposed as $ \psi(\mathbf{x}, t) = c(t) + \phi(\mathbf{x}, t) $, where $ \phi $ represents small fluctuations, i.e. $ \phi \ll c $. 
The condensate part evolves as $ c(t)=c_0 e^{i \rho_0 t} $, with $ \rho_0 = | \hat\psi(0, t)|^2 $.
Linearizing the system the new dispersion relation $ \omega(k)= \rho_0 \pm k \sqrt{k^2 + 2 \rho_0} $ can be found, known in the literature as Bogoliubov dispersion.  
In figure \ref{dispersion-hv} we present the numerical evaluation of the dispersion relation taken at the final stage of simulation, case {\bf (b)}; results show excellent agreement  with the theoretical Bogoliubov curve.
For very strong condensate, $ \rho_0 \gg k^2$, the Bogoliubov waves become acoustic and $\omega(k)= k \sqrt{2\rho_0} $ (in a reference frame rotating with the condensate speed $ \rho_0$).
Such acoustic WWT was considered in \cite{zakharov1970sat} and the respective KZ spectrum,  $E^{1D}(k) \sim k^{-3/2}$, is very close to Kolmogorov $-5/3$.
To make comparison with our results we have to take into account that, in this regime,  $ E^{1D} \sim \rho_0 \, n^{1D} $ \cite{zakharov2005dbe}.
By looking at the late time spectrum {\bf(b)} in figure \ref{spectrum-hv} we see that our results are consistent with the Zakharov-Sagdeev prediction $k^{-3/2}$ for the three-wave acoustic turbulence, although the scaling range is tiny and not well developed because of the long transient range with peaks \cite{falkovich1988efc}.
%To make a more definite conclusion one has to increase the inertial range.

\begin{figure}
\includegraphics[scale=0.24]{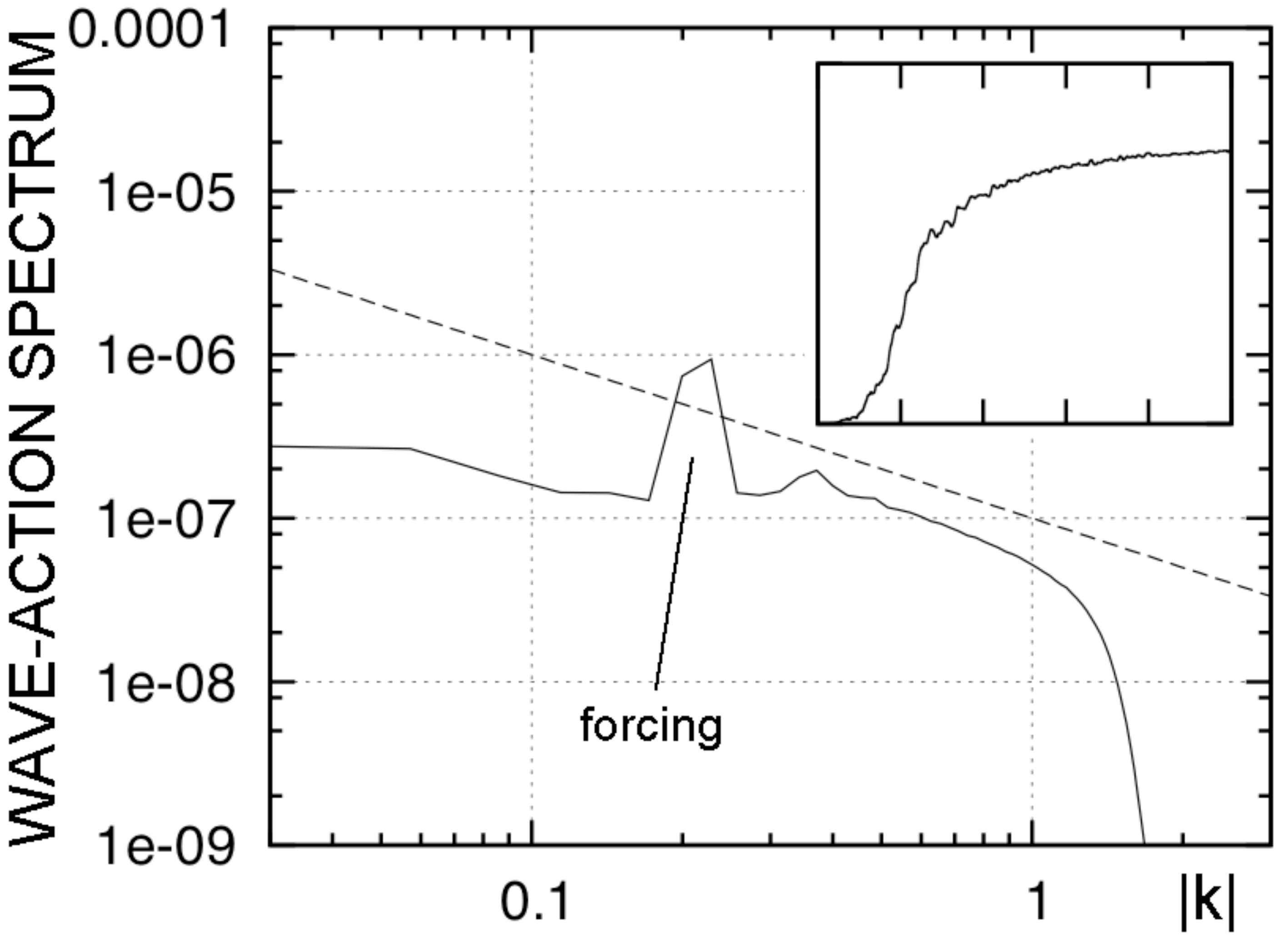}
\caption{Spectrum $ n^{1D}(k, t) $ at final stage of simulation in the presence of the friction term. 
The $ k^{-1} $ prediction of WWT is also shown. 
Inset: $ c_0=|\hat\psi(0,t)|$ as a function of time.\label{spectrum-hvf}}
\end{figure}

Note that the condensate keeps growing in this simulation. 
In order to avoid such growth, a dissipation can be included at wave numbers lower that the ones corresponding to forcing. 
Different options are available. 
Firstly, we will use a {\it friction}-type  dissipation which takes the form, in Fourier space, $ \hat{\mathcal{D}}= i \mu \theta(k^*-k) \hat \psi $, where $ \theta $ is the Heaviside step function, $ k^*=9 \Delta k $ corresponds to lowest wavenumber forced and $\mu$ is a friction coefficient which has been set to $ \mu=1 \times 10^{-4} $. 
We present our stationary state solution in figure \ref{spectrum-hvf}. 
The resulting spectral slope is consistent with the  prediction of the WWT theory. 
The growth of the condensate is now stopped by friction, as shown in the inset, and transition  from the four-wave to a three-wave regime is prevented.

\begin{figure}
\includegraphics[scale=0.24]{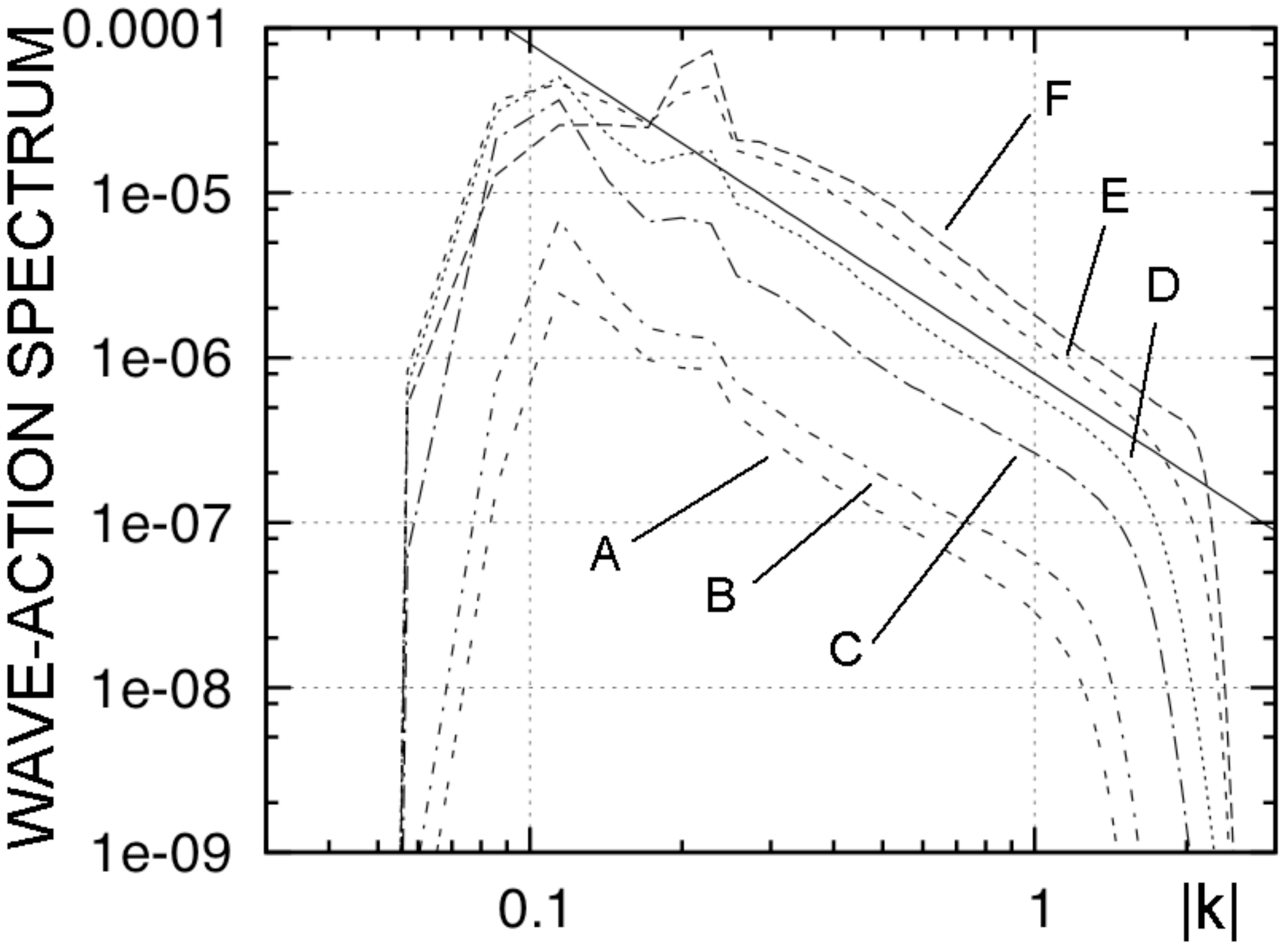}
\caption{Wave-action $ n^{1D}(k, t_f) $ spectrum at final stage of simulation with hypo-viscosity for different forcing coefficient: $ f_0=0.05 $ {\bf (A)}, $ f_0=0.1 $ {\bf (B)}, $ f_0=0.5 $ {\bf (C)}, $ f_0=1.0 $ {\bf (D)}, $ f_0=2 $ {\bf (E)}, $ f_0=3 $ {\bf (F)}.
A $ k^{-2} $ slope is also plotted.\label{spectrum-hlv}}
\end{figure}

Another common way of damping the low wave numbers consists, in analogy to what is done at high wave numbers, in including a {\it hypo-viscosity} term $ \mathcal{D}= i \nu_l(\nabla^{-2})^m \psi $ in equation (\ref{nls0}) and suppressing the condensate in Fourier space (mode $ \mathbf{k}=0 $). 
In our simulations, we have chosen $ \nu_l = 1 \times 10^{-18} $ and $ m=8 $.  
In figure \ref{spectrum-hlv} we show the stationary states achieved with this new damping term for different forcing coefficient $ f_0 $.  
The observed  spectrum is clearly much steeper than the WWT prediction and it is reasonably fitted by a power law $k^{-2}$ for forcing $ f_0 $ in a wide range  (two orders of magnitude). 
It seems that the direct energy cascade is strongly influenced by the accumulation of wave-action at wave numbers below the forcing. 
In other words, a  sharp dissipative term at low wave numbers can cause an {\itshape infrared bottleneck effect}.
Similar behavior (steeper spectrum) has been observed recently in numerical simulations for water waves \cite{korotkevich2008sns}.

\begin{figure}
\includegraphics[scale=0.24]{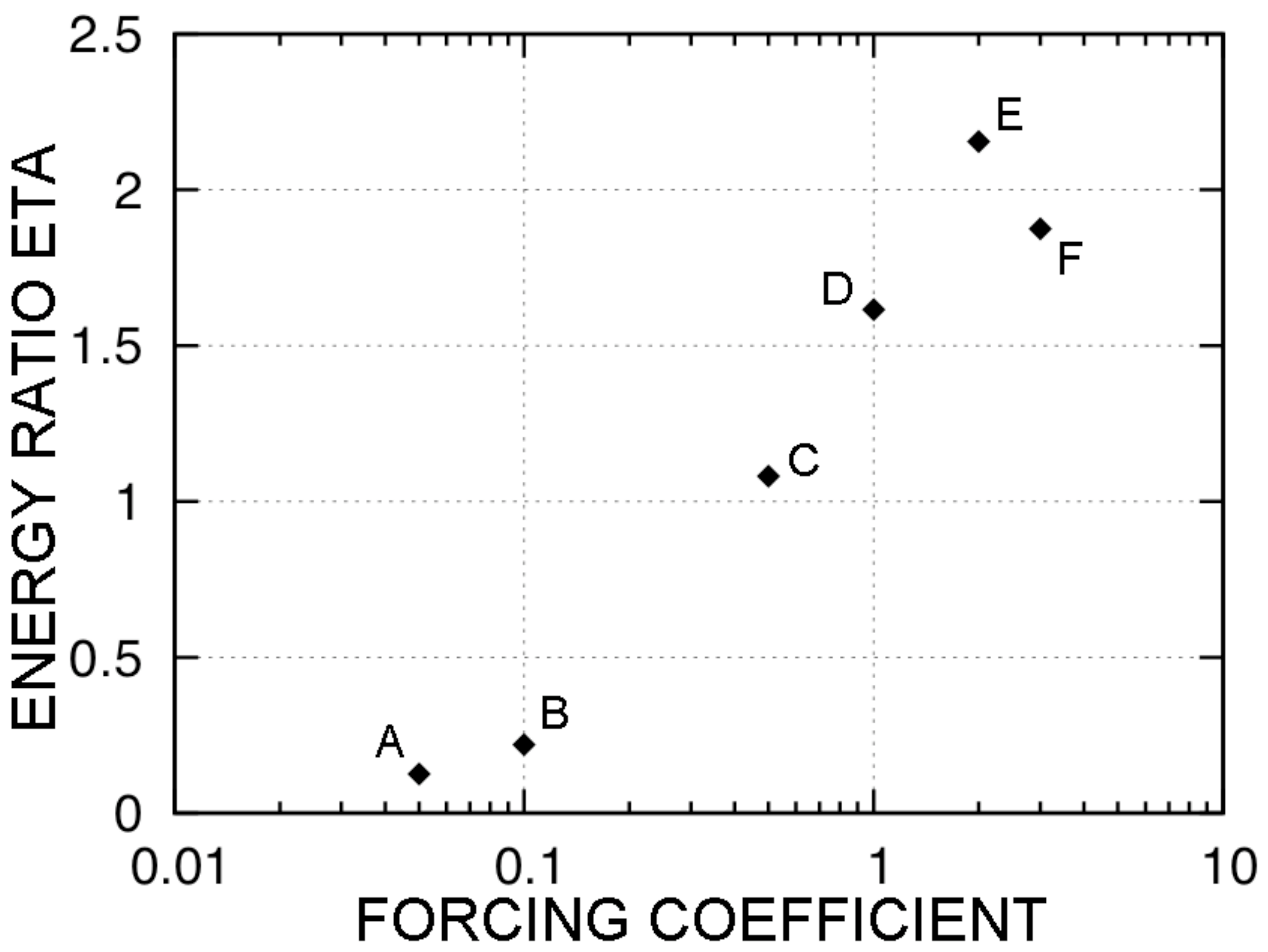}
\caption{Energy ratio $  \eta=H_{NL}/H_{LIN} $ evaluated at steady non-equilibrium state with the hypo-viscous dissipation for different forcing coefficient $ f_0 $ (see figure \ref{spectrum-hlv} and its label). \label{energy_ratio-hlv}}
\end{figure}

To understand these results we try to catch the level of nonlinearity in the system by considering the ratio $ \eta=H_{NL}/H_{LIN} $. 
Note that integral quantities are not always relevant because we are interested at $ \eta $ in the inertial range and both energies may be strongly influenced by what happens, for example, in the forcing or in the low wave number region. 
In the case where WWT prediction are confirmed (figure \ref{spectrum-hvf}), $ \eta \approx 1.06 $; apparently WWT condition is not valid in this case but probably most of the nonlinear energy, in Fourier space, is stacked at low wave-numbers and so, in the inertial range, the nonlinearity remains weak.
It is instructive to look now at $ \eta $ in simulations with hypo-viscosity that give the $ k^{-2} $ slope. 
As we can see in figure \ref{energy_ratio-hlv}, even by increasing the forcing coefficient $ f_0 $ by two order of magnitude, the ratio $ \eta $ remains of order one. 
In those cases it is reasonable to think that the infrared bottleneck accumulation lead to the growth of the nonlinear terms until they become comparable to  the linear ones in the inertial range.

This observations lead to a "critical balance" (CB) conjecture that the systems saturates in a state where the linear and the nonlinear timescales are balanced on a {\em scale-by-scale} basis.
The name CB is borrowed from MHD turbulence were it was originally proposed in \cite{goldreich1995tti}. 
Even though not called by this name, CB-like ideas were put forward in the past for several other physical wave systems, notably the water surface gravity waves where the CB condition leads to the famous Phillips spectrum.
Indeed, in the Phillips spectrum the wave steepness is saturated by the  wave breaking process  when a fluid particle cannot stay attached to the water surface because its downward acceleration becomes equal to the gravity constant, which occurs when the nonlinear time scale becomes of order of the linear one.

Now we propose a similar CB state may form in  GPE turbulence with hypo-viscous dissipation (or another kind of low-${\bf k}$ dissipation which is sharp enough to lead to the infrared bottleneck).
Namely,  when the low-${\bf k}$ range is over-dissipated by strong hypo-viscosity, the inverse cascade tendency tends to accumulate the spectrum  at low ${\bf k}$'s until the critical balance is reached and the spectrum is saturated. 
When the size of the nonlinear term, locally in Fourier space, becomes of the same order as the linear, which is the critical balance condition, the inverse cascade is arrested and the further (infrared) bottleneck accumulation is halted. 
One could also qualitatively view this as a set of nonlinear coherent structures, in this case solitons or/and vortices, whose amplitude is limited by the linear dispersion (i.e. stronger solitons would break into the weaker ones and incoherent waves).

We now present an estimate the CB spectrum in the GPE model. 
Equating the linear and the nonlinear terms for equation (\ref{nls0}) written in Fourier space, we have
\begin{equation}
k^2 |\hat{\psi}_k| \sim  |\hat{\psi}_k|^3 k^{6},
\label{cb1}
\end{equation}
which gives for the 1D wave action spectrum
\begin{equation}
n^{1D}(k) = 4\pi k^2 n(k) =4\pi k^2 (L/2\pi)^3 |\hat{\psi}_k|^2 \sim k^{-2}.
\label{cb2}
\end{equation}
Note that in evaluating the nonlinear term in (\ref{cb1}) we replaced each integration over $ \mathbf{k} $ with $ k^3 $ which implies that the modes with wave numbers $ \mathbf{k} $ are correlated with other $ k $-modes with wavenumbers of the similar values $ \sim \mathbf{k} $. 
This is consistent with our assumption that the nonlinearity is scale-by-scale of the same order as the linear term. 
%For comparison, in weakly nonlinear waves with random phases the nonlinear interactions would be weakened by statistical cancelations (like in the kinetic equation). 
The $ k^{-2} $ prediction in equation (\ref{cb2}) is consistent with our numerical simulations shown in figure \ref{spectrum-hlv}.

Concluding, we have performed numerical simulations of the 3D GPE with forcing and dissipation. 
%This system appears to be thornier than the 3D Navier-Stokes turbulence case, which has only a direct energy cascade. 
The direct energy cascade range is strongly influenced by the second conserved quantity, the wave-action $ N $, which has an inverse cascade tendency; results depend on how the low $ k $'s are damped. 
We have observed three different types of universal behavior roughly corresponding to situations where the largest scales are either non-dissipative, or damped by an efficient (e.g. friction-type) dissipation, or  or damped by an inefficient (e.g. hypo-viscosity) dissipation.
In the first case turbulence is not steady:  initial direct energy cascade, with a spectrum in good agreement with predictions of the WWT theory, is followed by condensation at the largest scales and
 a transition from a four to a three-wave interactions with a clearly Bogoliubov dispersion relation characteristic to this regime.
In the second case, the wave-action cascade is effectively absorbed so that there is no condensation, and we observe a steady state spectrum which is in good agreement with the WWT theory.
In the third case,  the dissipation  is not so efficient and an infrared bottleneck forms in the spectrum.
In this regime we observe a robust steady state spectrum which could be explained by a phenomenological "critical balance" proposition where the linear and nonlinear timescales are balanced on the scale-by-scale basis.

\bibliography{references_final}

\begin{thebibliography}{21}
\expandafter\ifx\csname natexlab\endcsname\relax\def\natexlab#1{#1}\fi
\expandafter\ifx\csname bibnamefont\endcsname\relax
  \def\bibnamefont#1{#1}\fi
\expandafter\ifx\csname bibfnamefont\endcsname\relax
  \def\bibfnamefont#1{#1}\fi
\expandafter\ifx\csname citenamefont\endcsname\relax
  \def\citenamefont#1{#1}\fi
\expandafter\ifx\csname url\endcsname\relax
  \def\url#1{\texttt{#1}}\fi
\expandafter\ifx\csname urlprefix\endcsname\relax\def\urlprefix{URL }\fi
\providecommand{\bibinfo}[2]{#2}
\providecommand{\eprint}[2][]{\url{#2}}

\bibitem[{\citenamefont{Anderson et~al.}(1995)\citenamefont{Anderson, Ensher,
  Matthews, Wieman, and Cornell}}]{anderson1995obe}
\bibinfo{author}{\bibfnamefont{M.}~\bibnamefont{Anderson}},
  \bibinfo{author}{\bibfnamefont{J.}~\bibnamefont{Ensher}},
  \bibinfo{author}{\bibfnamefont{M.}~\bibnamefont{Matthews}},
  \bibinfo{author}{\bibfnamefont{C.}~\bibnamefont{Wieman}}, \bibnamefont{and}
  \bibinfo{author}{\bibfnamefont{E.}~\bibnamefont{Cornell}},
  \bibinfo{journal}{Science} \textbf{\bibinfo{volume}{269}},
  \bibinfo{pages}{198} (\bibinfo{year}{1995}).

\bibitem[{\citenamefont{Davis et~al.}(1995)\citenamefont{Davis, Mewes, Andrews,
  Van~Druten, Durfee, Kurn, and Ketterle}}]{davis1995bec}
\bibinfo{author}{\bibfnamefont{K.}~\bibnamefont{Davis}},
  \bibinfo{author}{\bibfnamefont{M.}~\bibnamefont{Mewes}},
  \bibinfo{author}{\bibfnamefont{M.}~\bibnamefont{Andrews}},
  \bibinfo{author}{\bibfnamefont{N.}~\bibnamefont{Van~Druten}},
  \bibinfo{author}{\bibfnamefont{D.}~\bibnamefont{Durfee}},
  \bibinfo{author}{\bibfnamefont{D.}~\bibnamefont{Kurn}}, \bibnamefont{and}
  \bibinfo{author}{\bibfnamefont{W.}~\bibnamefont{Ketterle}},
  \bibinfo{journal}{Physical Review Letters} \textbf{\bibinfo{volume}{75}},
  \bibinfo{pages}{3969} (\bibinfo{year}{1995}).

\bibitem[{\citenamefont{Bose}(1924)}]{bose1924pgl}
\bibinfo{author}{\bibfnamefont{S.}~\bibnamefont{Bose}},
  \bibinfo{journal}{Zeitschrift fur Physik} \textbf{\bibinfo{volume}{26}},
  \bibinfo{pages}{178} (\bibinfo{year}{1924}).

\bibitem[{\citenamefont{Einstein}(1925)}]{einstein1925paw}
\bibinfo{author}{\bibfnamefont{A.}~\bibnamefont{Einstein}},
  \bibinfo{journal}{Klasse, Sitzungsberichte} \textbf{\bibinfo{volume}{23}}
  (\bibinfo{year}{1925}).

\bibitem[{\citenamefont{Pitaevskii and Stringari}(2003)}]{pitaevskii2003bec}
\bibinfo{author}{\bibfnamefont{L.}~\bibnamefont{Pitaevskii}} \bibnamefont{and}
  \bibinfo{author}{\bibfnamefont{S.}~\bibnamefont{Stringari}},
  \emph{\bibinfo{title}{{Bose-Einstein condensation}}}
  (\bibinfo{publisher}{Oxford University Press, USA}, \bibinfo{year}{2003}).

\bibitem[{\citenamefont{Dyachenko et~al.}(1992)\citenamefont{Dyachenko, Newell,
  Pushkarev, and Zakharov}}]{dyachenko1992wtc}
\bibinfo{author}{\bibfnamefont{S.}~\bibnamefont{Dyachenko}},
  \bibinfo{author}{\bibfnamefont{A.~C.} \bibnamefont{Newell}},
  \bibinfo{author}{\bibfnamefont{A.}~\bibnamefont{Pushkarev}},
  \bibnamefont{and} \bibinfo{author}{\bibfnamefont{V.~E.}
  \bibnamefont{Zakharov}}, \bibinfo{journal}{Physica D: Nonlinear Phenomena}
  \textbf{\bibinfo{volume}{57}}, \bibinfo{pages}{96} (\bibinfo{year}{1992}).

\bibitem[{\citenamefont{Berloff and Svistunov}(2002)}]{berloff2002ssn}
\bibinfo{author}{\bibfnamefont{N.}~\bibnamefont{Berloff}} \bibnamefont{and}
  \bibinfo{author}{\bibfnamefont{B.}~\bibnamefont{Svistunov}},
  \bibinfo{journal}{Physical Review A} \textbf{\bibinfo{volume}{66}},
  \bibinfo{pages}{13603} (\bibinfo{year}{2002}).

\bibitem[{\citenamefont{Nazarenko and Onorato}(2006)}]{nazarenko2006wta}
\bibinfo{author}{\bibfnamefont{S.}~\bibnamefont{Nazarenko}} \bibnamefont{and}
  \bibinfo{author}{\bibfnamefont{M.}~\bibnamefont{Onorato}},
  \bibinfo{journal}{Physica D: Nonlinear Phenomena}
  \textbf{\bibinfo{volume}{219}}, \bibinfo{pages}{1} (\bibinfo{year}{2006}).

\bibitem[{\citenamefont{Koplik and Levine}(1993)}]{koplik1993vrs}
\bibinfo{author}{\bibfnamefont{J.}~\bibnamefont{Koplik}} \bibnamefont{and}
  \bibinfo{author}{\bibfnamefont{H.}~\bibnamefont{Levine}},
  \bibinfo{journal}{Physical Review Letters} \textbf{\bibinfo{volume}{71}},
  \bibinfo{pages}{1375} (\bibinfo{year}{1993}).

\bibitem[{\citenamefont{Nore et~al.}(1997)\citenamefont{Nore, Abid, and
  Brachet}}]{nore1997dkt}
\bibinfo{author}{\bibfnamefont{C.}~\bibnamefont{Nore}},
  \bibinfo{author}{\bibfnamefont{M.}~\bibnamefont{Abid}}, \bibnamefont{and}
  \bibinfo{author}{\bibfnamefont{M.}~\bibnamefont{Brachet}},
  \bibinfo{journal}{Physics of Fluids} \textbf{\bibinfo{volume}{9}},
  \bibinfo{pages}{2644} (\bibinfo{year}{1997}).

\bibitem[{\citenamefont{Abid et~al.}(2003)\citenamefont{Abid, Huepe, Metens,
  Nore, Pham, Tuckerman, and Brachet}}]{abid2003gpd}
\bibinfo{author}{\bibfnamefont{M.}~\bibnamefont{Abid}},
  \bibinfo{author}{\bibfnamefont{C.}~\bibnamefont{Huepe}},
  \bibinfo{author}{\bibfnamefont{S.}~\bibnamefont{Metens}},
  \bibinfo{author}{\bibfnamefont{C.}~\bibnamefont{Nore}},
  \bibinfo{author}{\bibfnamefont{C.}~\bibnamefont{Pham}},
  \bibinfo{author}{\bibfnamefont{L.}~\bibnamefont{Tuckerman}},
  \bibnamefont{and} \bibinfo{author}{\bibfnamefont{M.}~\bibnamefont{Brachet}},
  \bibinfo{journal}{Fluid Dynamics Research} \textbf{\bibinfo{volume}{33}},
  \bibinfo{pages}{509} (\bibinfo{year}{2003}).

\bibitem[{\citenamefont{Parker and Adams}(2005)}]{parker2005edt}
\bibinfo{author}{\bibfnamefont{N.}~\bibnamefont{Parker}} \bibnamefont{and}
  \bibinfo{author}{\bibfnamefont{C.}~\bibnamefont{Adams}},
  \bibinfo{journal}{Physical Review Letters} \textbf{\bibinfo{volume}{95}},
  \bibinfo{pages}{145301} (\bibinfo{year}{2005}).

\bibitem[{\citenamefont{Kobayashi and Tsubota}(2005)}]{kobayashi2005kss}
\bibinfo{author}{\bibfnamefont{M.}~\bibnamefont{Kobayashi}} \bibnamefont{and}
  \bibinfo{author}{\bibfnamefont{M.}~\bibnamefont{Tsubota}},
  \bibinfo{journal}{Physical Review Letters} \textbf{\bibinfo{volume}{94}},
  \bibinfo{pages}{65302} (\bibinfo{year}{2005}).

\bibitem[{\citenamefont{Kobayashi and Tsubota}(2007)}]{kobayashi2007qtt}
\bibinfo{author}{\bibfnamefont{M.}~\bibnamefont{Kobayashi}} \bibnamefont{and}
  \bibinfo{author}{\bibfnamefont{M.}~\bibnamefont{Tsubota}},
  \bibinfo{journal}{Physical Review A} \textbf{\bibinfo{volume}{76}},
  \bibinfo{pages}{045603} (\bibinfo{year}{2007}).

\bibitem[{\citenamefont{Zakharov et~al.}(1992)\citenamefont{Zakharov, L'vov,
  and Falkovich}}]{zakharov41kst}
\bibinfo{author}{\bibfnamefont{V.}~\bibnamefont{Zakharov}},
  \bibinfo{author}{\bibfnamefont{V.}~\bibnamefont{L'vov}}, \bibnamefont{and}
  \bibinfo{author}{\bibnamefont{Falkovich}}, \emph{\bibinfo{title}{{Kolmogorov
  Spectra of Turbulence 1: Wave Turbulence}}}
  (\bibinfo{publisher}{Springer-Verlag}, \bibinfo{year}{1992}).

\bibitem[{\citenamefont{Zakharov et~al.}(1985)\citenamefont{Zakharov, Musher,
  and Rubenchik}}]{zakharov1985had}
\bibinfo{author}{\bibfnamefont{V.}~\bibnamefont{Zakharov}},
  \bibinfo{author}{\bibfnamefont{S.}~\bibnamefont{Musher}}, \bibnamefont{and}
  \bibinfo{author}{\bibfnamefont{A.}~\bibnamefont{Rubenchik}},
  \bibinfo{journal}{Physics Reports} \textbf{\bibinfo{volume}{129}}
  (\bibinfo{year}{1985}).

\bibitem[{\citenamefont{Falkovich and Shafarenko}(1988)}]{falkovich1988efc}
\bibinfo{author}{\bibfnamefont{G.}~\bibnamefont{Falkovich}} \bibnamefont{and}
  \bibinfo{author}{\bibfnamefont{A.}~\bibnamefont{Shafarenko}},
  \bibinfo{journal}{Soviet Physics - JETP} \textbf{\bibinfo{volume}{68}},
  \bibinfo{pages}{1393} (\bibinfo{year}{1988}).

\bibitem[{\citenamefont{Zakharov and Nazarenko}(2005)}]{zakharov2005dbe}
\bibinfo{author}{\bibfnamefont{V.}~\bibnamefont{Zakharov}} \bibnamefont{and}
  \bibinfo{author}{\bibfnamefont{S.}~\bibnamefont{Nazarenko}},
  \bibinfo{journal}{Physica D: Nonlinear Phenomena}
  \textbf{\bibinfo{volume}{201}}, \bibinfo{pages}{203} (\bibinfo{year}{2005}).

\bibitem[{\citenamefont{Zakharov and Sagdeev}(1970)}]{zakharov1970sat}
\bibinfo{author}{\bibfnamefont{V.}~\bibnamefont{Zakharov}} \bibnamefont{and}
  \bibinfo{author}{\bibfnamefont{R.}~\bibnamefont{Sagdeev}},
  \bibinfo{journal}{Soviet Physics - Doklady} \textbf{\bibinfo{volume}{15}}
  (\bibinfo{year}{1970}).

\bibitem[{\citenamefont{Korotkevich}(2008)}]{korotkevich2008sns}
\bibinfo{author}{\bibfnamefont{A.}~\bibnamefont{Korotkevich}},
  \bibinfo{journal}{Physical Review Letters} \textbf{\bibinfo{volume}{101}},
  \bibinfo{pages}{074504} (\bibinfo{year}{2008}).

\bibitem[{\citenamefont{Goldreich and Sridhar}(1995)}]{goldreich1995tti}
\bibinfo{author}{\bibfnamefont{P.}~\bibnamefont{Goldreich}} \bibnamefont{and}
  \bibinfo{author}{\bibfnamefont{S.}~\bibnamefont{Sridhar}},
  \bibinfo{journal}{The Astrophysical Journal} \textbf{\bibinfo{volume}{438}},
  \bibinfo{pages}{763} (\bibinfo{year}{1995}).

\end{thebibliography}

\end{document}